\documentclass[prb,amsmath,amssymb,twocolumn,superscriptaddress ]{revtex4-1}
\usepackage{graphicx}
\usepackage{epsfig}
\usepackage{dcolumn}
\usepackage{bm}
\usepackage{rotating}
\usepackage[table]{xcolor}
\usepackage[english]{babel}

\begin{document}
\widetext 

\title{Phonon-assisted damping of plasmons in three- and two-dimensional metals}

\author{Fabio Caruso}
\affiliation{Institut f\"ur Physik and IRIS Adlershof, Humboldt-Universit\"at zu Berlin, Berlin, Germany}
\author{Dino Novko}
\affiliation{Institut f\"ur Chemie und Biochemie, Freie Universit\"at 
Berlin, Takustr. 3, 14195 Berlin, Germany}
\author{Claudia Draxl}
\affiliation{Institut f\"ur Physik and IRIS Adlershof, Humboldt-Universit\"at zu Berlin, Berlin, Germany}
\date{\today}
\pacs{}

\date{\today}
\begin{abstract}
We investigate the effects of crystal lattice vibrations on the dispersion of plasmons. 
{The loss function of the homogeneous 
electron gas (HEG) in two and three dimensions is evaluated numerically }
in presence of electronic coupling to an optical phonon mode. 
{Our} calculations are based on many-body perturbation 
theory for the dielectric function as formulated by 
the Hedin-Baym equations in the Fan-Migdal approximation. 
{The coupling to phonons broadens the spectral signatures of plasmons 
in the electron-energy loss spectrum (EELS) and it induces the decay of 
plasmons on timescales shorter than 1~ps.
Our results further reveal the formation of a kink in 
the plasmon dispersion of the 2D HEG, which marks the 
onset of plasmon-phonon scattering.
Overall, these features constitute a fingerprint 
of plasmon-phonon coupling in the EELS of simple metals.}
{It is shown} that these effects may 
be accounted for by resorting to a simplified treatment of the 
electron-phonon interaction which is amenable to first-principles calculations. 
\end{abstract}

\keywords{}
\maketitle

\section{Introduction}

The coupling between electrons and crystal lattice vibrations (phonons)
affects pervasively the quantum behaviour of condensed matter.\cite{GiustinoRMP} 
It underpins a broad spectrum of emergent phenomena, 
including kinks and polaronic satellites in photoemission,\cite{Damascelli2003,Moser2013,Baumberger2016,Verdi2017,King2017} 
the temperature dependence of electronic bands,\cite{Logothetidis1992,Eiguren2008,Eiguren2009,GiustinoPRL2010,Ponce2014,Antonius2014} 
and optical spectra,\cite{Jellison1983,Lautenschlager1987,Marini2008}
(non-adiabatic) Kohn anomalies,\cite{Kohn1959,Piscanec2004,Lazzeri2006,CarusoPRL2017} 
and conventional superconductivity.\cite{schrieffer1983theory} 
Electron-phonon scattering also constitutes one of the primary 
decay channels that drives dissipation phenomena in solids such as,
e.g., hot carrier relaxation,\cite{bernardi/2014,Bernardi2015,CarusoPRB2016} 
the decay of collective charge-density fluctuations (plasmons),\cite{BernardiSPP,Atwater2016,Novko2017} 
and phonon damping.\cite{grimvall1981electron}
A quantitative understanding of these mechanisms through first-principles 
simulations may open new ways towards the identification of materials for applications 
where the characteristic time-scales of the dissipation processes
are critical for the device function. 

Plasmonics, for instance, employs plasmons and  surface plasmon polaritons (SPP) to manipulate 
light-matter interactions at the nano-scale, and it 
has proven successful in an interdisciplinary 
area of applications, encompassing, e.g., photovoltaics,\cite{Knight2011} radiation therapy,\cite{Hirsch2003,ONeal2004} 
and sensing.\cite{Brongersma}
Since long plasmon lifetimes precondition the functionality of plasmonic devices,\cite{Khurgin} 
dissipation mechanisms arising from electron-electron and electron-lattice
scattering, which are at the origin of plasmon damping (the decay of plasmons and SPP), 
presently constitute a major bottleneck for plasmonics. 
The development of predictive theories and simulation 
tools for the study of phonon-assisted dissipation may 
provide a new impulse to unravel the origin of plasmon 
damping and address the issue of quantum losses in plasmonics.

A plasmon is excited at momenta ${\bf q}$ and 
frequencies $\omega_{\rm pl}$ which correspond 
to zeros of the macroscopic dielectric function.\cite{pines1999elementary} 
A fully quantum-mechanical description of phonon-assisted plasmon damping, 
where a plasmon annihilates through the emission of a phonon, 
requires to account for the interplay between plasmons and phonons
by generalizing the dielectric function to the coupled electron-phonon system.\cite{Holstein1964}
Recent years have witnessed remarkable advancements along this line. 
The temperature dependence of the absorption onset of
indirect band-gap semiconductors, a phenomenon dominated
by phonon-assisted processes, has motivated the formulation
of a theory of optical absorption based on Fermi's golden
rule,\cite{Noffsinger2012}  as well as a generalization
of William-Lax theory for phonon-assisted optical absorption 
to first-principles calculations\cite{Patrick2014,Zacharias2015}
and its extension to non-stochastic one-shot calculations.\cite{Zacharias2016}
An alternative procedure to account for the coupling to phonons 
in the evaluation of the dielectric function  
consists in introducing quasiparticle effects due to electron-phonon interaction 
in the Bethe-Salpeter equation, a procedure that has been applied to silicon\cite{Marini2008}
and to transition metal dichalcogenides,\cite{Molina2016} and through the formulation 
of a many-body theory of exciton-phonon interaction.\cite{Antonious2017}
While these works have primarily addressed the temperature
dependence of absorption spectroscopy experiments, only recently 
the effects of electron-phonon coupling on the damping of plasmons 
have been accounted for from first principles.\cite{Novko2017} 

{
In this manuscript, we introduce a field-theoretic formalism which,
based on the Hedin-Baym equations, 
a formally exact formulation of the problems of interacting
electrons and phonons in the harmonic approximation,
allows the investigation of the damping of plasmons induced by the electron-phonon interaction.
Numerical results for the plasmon dispersion of the homogeneous
electron gas (HEG) are reported in two (2D) and three dimensions (3D) at
carrier densities representative of metals and highly-doped semiconductors.
Plasmon-damping effects arising from electronic coupling to a non-dispersive 
optical phonon are investigated through the computation of the loss function. 
In 3D, the electron-phonon interaction introduces finite-lifetime effects that
significantly broaden the plasmon dispersion, providing
a fingerprint of phonon-assisted plasmon damping in 3D metals.
In 2D metallic systems, the proximity
of the phonon and plasmon energies at long 
wavelengths gives rise to a kink in the plasmon dispersion.
This novel spectroscopic signature of plasmon-phonon coupling,
reminiscent of the photoemission kinks of superconductors\cite{lanzara/2001} and
pristine graphene,\cite{Botswick} manifests itself at the energy and momentum that mark the
onset of plasmon-phonon scattering.
The accuracy of approximate techniques to account for plasmon-phonon
interaction in first-principles calculations at a reduced computational
cost is further discussed.
Beside providing a many-body formalism for the description of
phonon-assisted plasmon damping, this study contributes to unveil novel spectral
signatures resulting from  plasmon-phonon coupling, a result that may prove useful
for the characterization of plasmonic materials with reduced dimensionality such
as, e.g., doped graphene or single-layer transition metal dichalcogenides.}

The manuscript is organized as follows. 
In Sec.~\ref{sec:theory} we report the Hedin-Baym equations for the 
dielectric function and the procedure followed for their solution 
in the Fan-Migdal approximation. 
Numerical results for the 3D HEG are presented in Sec.~\ref{sec:3D}, 
whereas Sec.~\ref{sec:2D0} illustrates the two-dimensional case. 
{In Sec.~\ref{sec:mem} the accuracy of approximate procedures to account
for electron-phonon coupling effects in the evaluation of the dielectric 
function are discussed. }
Finally, concluding remarks are presented in Sec.~\ref{sec:conc}. 

\section{Hedin-Baym equations for the dielectric function}\label{sec:theory}

{ Many-body perturbation theory provides a suitable framework to 
investigate the effects of the electron-phonon interaction on 
the plasmon dispersions of metals. 
Hedin's equations for the single-particle Green function 
constitute an exact formulation of the many-body problem 
under the assumption that the nuclei are fixed at their equilibrium 
position ({\it clamped-nuclei approximation}).\cite{Hedin1969}
Hedin's equations can be generalized to account for the coupling 
between nuclear and electronic degrees of freedom by explicitly considering 
the variation of the total electronic potential up to the second order in 
the atomic displacement (harmonic approximation). 
The resulting set of equations, denoted as the Hedin-Baym equations, 
constitute a formally exact framework for describing a 
electron-phonon system within the harmonic approximation.\cite{GiustinoRMP}
Within this formalism, the dielectric function is determined by the following 
set of equations: 
}
\begin{align}
\varepsilon(12) &= \delta(12) - \int d(3) v(13) P(32)  \nonumber \\  
P(12) & = - i \int d (34) G(13)G(41^+) \Gamma(342)  \nonumber  \\
G(12) &= G_0(12) + \int d(34) G_0(13)\Sigma(34)G(42)  \nonumber \\ 
\Sigma(12) &= i\hbar \int d(34) G(13) \Gamma (342) [W_e (41^+) + W_{\rm ph}(41^+) ]  \nonumber  
\end{align}
Here the collective index $1\equiv {\bf r}_1,t_1$ has been introduced. 
$\varepsilon$ is the dielectric function, $P$ the irreducible electronic polarization, 
$\Sigma$ the electron self-energy due to the combined effects of electrons and phonons, 
$G$ and $G_0$ the interacting and non-interacting Green functions, and 
$\Gamma$ the three-point vertex function. 
$v$ denotes the bare Coulomb interaction, whereas 
$W_{\rm e}$ and $W_{\rm ph}$ are the screened Coulomb interactions due to 
electron-electron and electron-phonon interaction, respectively, which are 
related by the expression\cite{GiustinoRMP} (in symbolic notation) 
$W_{\rm ph} = W_{\rm e} D W_{\rm e}$, with $D$ being the phonon Green function. 

{Analogously to the $GW$ method,\cite{Hedin1969,hybertsenlouie1986}
the Hedin-Baym equations should be solved 
self-consistently\cite{holmvonbarth1998,PhysRevB.88.075105,CarusoJCTC} 
owing to the interdependence of polarization, 
Green function, and self-energy.
In the following, to define a practical procedure for the computational evaluation of 
the dielectric function, we restrict ourselves to consider 
one-shot calculations of the polarizability and dielectric function, 
where the Green function is obtained from a single solution of the 
Dyson equation.
Additionally,  we set the vertex function to 
$\Gamma(123) = \delta(13)\delta(23)$, which corresponds 
to the Fan-Migdal approximation for the self-energy.
In practice, this approximation consists
in ignoring the contribution of electron-phonon coupling 
on the interaction between electrons and holes. 
}
We concentrate on the electron-phonon interaction, and thus ignore 
the effects of the electron-electron interaction beyond the 
single-particle picture by neglecting the electronic part of 
the screened interaction $W_e$ in the self-energy $\Sigma$. 
Within these approximations, and considering the Fourier
transform of the Hedin-Baym equations to momentum and frequency, 
the defining equation for the dielectric function of the coupled 
electron-phonon system may be rewritten as:
%
\begin{figure}
   \includegraphics[width=0.35\textwidth]{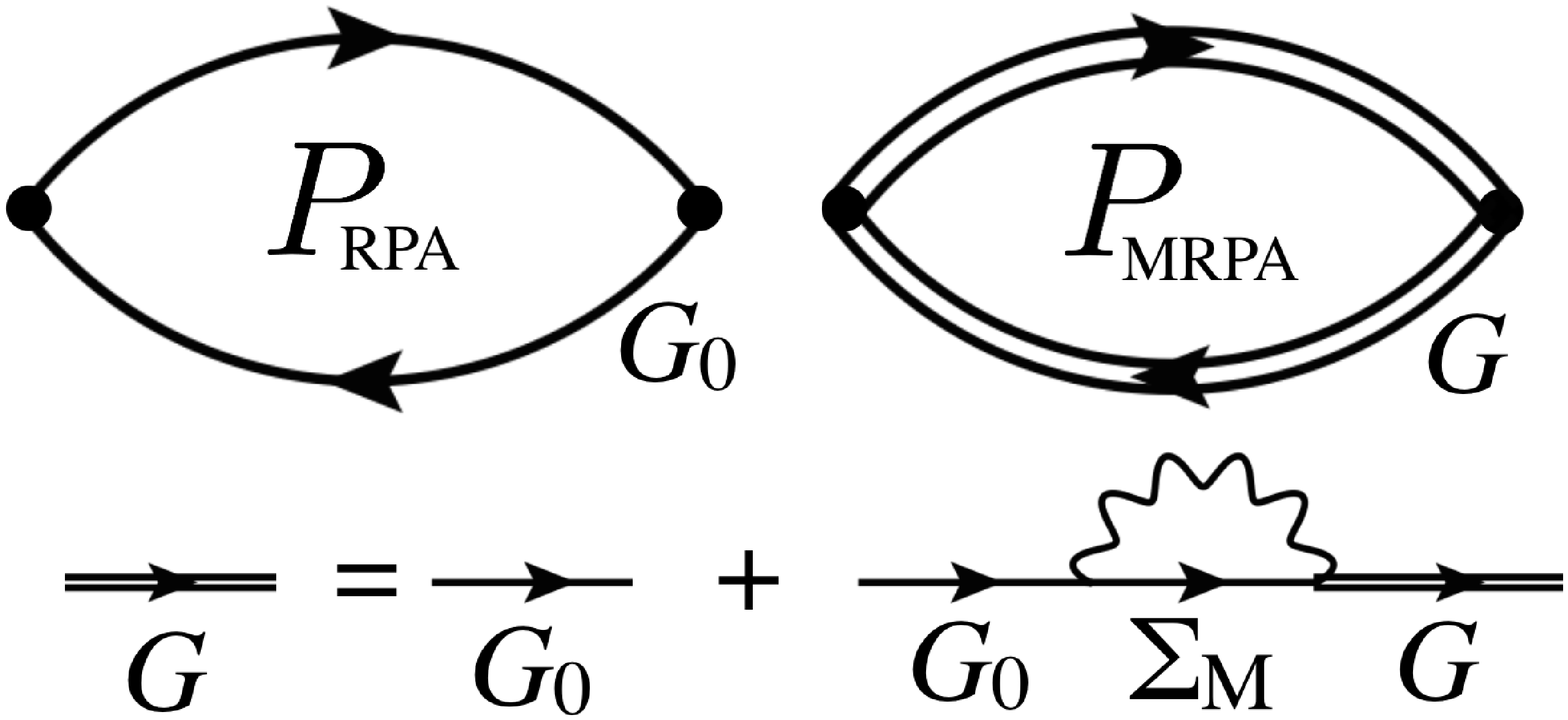} 
\caption{{Feynman diagrams for the irreducible polarization:
Left: RPA, right: polarizability obtained through the inclusion of 
electron-phonon coupling effects at first-order within the Fan-Migdal
approximation (MRPA).}
}
\label{figrpa}
\end{figure}
%
\begin{align}
\varepsilon({\bf q},\omega) &= 1-v({\bf q})P({\bf q},\omega) \label{eq:eps1} \\
P({\bf q},\omega) &= i \int \frac{d{\bf k}\,d\omega'}{(2\pi)^4} G({\bf q}+{\bf k},\omega+\omega') G({\bf k},\omega') \label{eq:chi1}\\
G({\bf k},\omega) &= [\hbar\omega - \epsilon_{\bf k} - \Sigma_{\rm M}({\bf k},\omega)]^{-1} \label{eq:g1}\\
\Sigma_{\rm M}({\bf k},\omega) &= \frac{i}{\hbar} \sum_\nu 
\int \frac{d{\bf q} d\omega'}{(2\pi)^4 } 
| g^\nu_{{\bf kq}}|^2 \times \nonumber \\ 
& \,\,\,\,\,\,\,\,\,\,\,\,\,\,\,\times G_0({\bf q}+{\bf k},\omega+\omega') D^\nu_0({\bf q},\omega')  \label{eq:sigma1}
\end{align}
where  $\epsilon_{\bf k}$ are the single-particle energies of non-interacting electrons, 
$g^\nu_{{\bf kq}}$ the electron-phonon coupling matrix elements,
$\nu$ and $\hbar\omega_{{\bf q}\nu}$ the phonon-mode index and energy, 
and $D^\nu_0({\bf q},\omega) = 2 \omega_{{\bf q}\nu} [ \omega^2 - \omega_{{\bf q}\nu}^2]^{-1}$ 
the non-interacting Green function for the $\nu$-th phonon.
$\Sigma_{\rm M}$ is the Fan-Migdal electron self-energy due to electron-phonon interaction, 
and it is derived following standard field-theoretic approaches.\cite{GiustinoRMP}
The procedure defined by Eqs.~(\ref{eq:eps1})-(\ref{eq:sigma1})
extends the random-phase approximation (RPA) for the irreducible 
polarization $P$ to the interacting electron-phonon system 
by replacing the non-interacting Green function with
a propagator {\it dressed} by the electron-phonon interaction in 
the Fan-Migdal approximation. 
A Feynman-diagrammatic representation of the polarizability in 
this approximation is compared to the 
RPA in Fig.~\ref{figrpa}. 
If the Fan-Migdal self-energy is set to zero, $G$ reduces 
to the non-interacting Green function $G_0$ and $P$ to the RPA. 

One may promptly verify that the combination of the Dyson equation with Eq.~(\ref{eq:chi1})
yields the following expression for the polarizability:\cite{kupcic2015}
\begin{align}
P({\bf q},\omega) = i \sum_{{\bf k}\sigma}\int \frac{d\omega'}{2\pi} 
\frac{ G({\bf k},\omega') - G({\bf k}+{\bf q},\omega + \omega') } 
{\hbar\omega + \epsilon_{\bf k} -\epsilon_{\bf k+q} + \Delta\Sigma({\bf k},{\bf q},\omega,\omega')}\nonumber
\end{align}
where the quantity 
$\Delta\Sigma({\bf k},{\bf q},\omega,\omega') = \Sigma_{\rm M}({\bf k},\omega') - \Sigma_{\rm M}({\bf k+q},\omega+\omega')$, sometimes denoted as 
electron-hole self-energy\cite{kupcic2015} or optical self-energy,\cite{Hwang2012} has been introduced.
The polarizability can thus be rewritten as:\cite{Holstein1964,kupcic2015}
\begin{align}\label{eq:chi0ph2}
P({\bf q},\omega) =  2\sum_{\bf k} \frac{ f_{\bf k} -f_{\bf k+q} } 
{\hbar\omega + \epsilon_{\bf k} -\epsilon_{\bf k+q} + \Delta\Sigma({\bf k},{\bf q},\omega,\epsilon_{\bf k}/\hbar)}.
\end{align}
Equation~\eqref{eq:chi0ph2} is obtained by neglecting 
the $\omega'$ dependence of the electron-hole self-energy
and performing the frequency integration only in the numerator, 
which yield the Fermi-Dirac occupation factors $f_{\bf k}$.
The combination of Eqs.~(\ref{eq:eps1}) and (\ref{eq:chi0ph2}) provides a 
practical recipe for incorporating the effects of electron-phonon coupling in the dielectric function.
Inspection of Eq.~(\ref{eq:chi0ph2}) suggests  that the interaction 
with phonons in the Fan-Migdal approximation may be approximately accounted for by replacing 
$\epsilon_{\bf k} \rightarrow \epsilon_{\bf k} + \Sigma_{\rm M}({\bf k}, \epsilon_{\bf k}/\hbar)$
and  $\epsilon_{\bf k+q} \rightarrow \epsilon_{\bf k+q} + \Sigma_{\rm M}({\bf k+q}, \epsilon_{\bf k+q}/\hbar + \omega) $ 
in the construction of the RPA polarizability. 

Generally, the inclusion of dynamical self-energy effects is expected to 
influence single-particle states by (i) renormalizing their energy due 
to the interaction with phonons; (ii) introducing an imaginary energy component 
that reflects the emergence of finite lifetime effects; 
(iii) leading to non-trivial dynamical (frequency-dependent) structures in the spectral function.
In the context of photoemission spectroscopy, 
the aspects (i-iii) have been thoroughly characterized via 
first-principles calculations and experiments, and they are known to 
underpin the formation of photoemission kinks,\cite{Damascelli2003}
the emergence of polaronic satellites 
close to the band edges,\cite{Moser2013,Baumberger2016,Verdi2017}
as well as the temperature-dependence of the band gap 
of insulators and semiconductors.\cite{Logothetidis1992,GiustinoPRL2010}
In the following, we compute the 
dielectric function  in the Fan-Migdal approximation to 
investigate the influence of (i-iii) on plasmons and electron-hole excitations. 

In the clumped-nuclei approximation, whereby electron-phonon
coupling is ignored, the inclusion of dynamical effects through 
the $GW$ self-energy in the evaluation of the dielectric function 
may result in an unphysical renormalization of the intensity of
calculated optical spectra if vertex corrections are neglected.\cite{DelSole1996}
For the case in which only the electron-electron interaction is
treated, vertex corrections compensate the dynamical self-energy 
effects and are important to achieve good agreement with experiments.\cite{Bechstedt1997,Marini2003} 
In the case of electron-phonon coupling, on the other hand, Migdal's 
theorem establishes that vertex corrections scale as $(m/M)^{1/2}$, 
where $m$ and $M$ are the electronic and ionic masses, respectively, 
suggesting that compensation effects due to the vertex should not alter
the optical properties significantly.\cite{Migdal1958}
However, the importance of vertex corrections for electron-phonon 
interaction remains an open question that calls for further 
investigations.

\section{Plasmon-phonon coupling in three dimensions}\label{sec:3D}

We account for the effects of electron-phonon interaction in the dielectric function 
through the numerical solution of Eqs.~\eqref{eq:eps1}-\eqref{eq:chi0ph2}.
We consider first a three-dimensional (3D) homogeneous electron gas with energy-momentum 
dispersion relations $\epsilon_{\bf k} = \frac{\hbar^2 k^2}{2 m^*}$,
where $m^*$ is the effective mass. {In the following 
we set $m^*=0.2\,m_{\rm e}$, a representative value for 
describing carriers at the band edges of doped semiconductors.}
The dielectric function in Eq.~\eqref{eq:eps1} is evaluated using the 
3D Fourier transform of the bare Coulomb interaction: $v({\bf q}) = 4\pi e^2/\varepsilon_0{\bf q}^2$, 
where $\varepsilon_0$ is the vacuum permittivity.
In absence of electron-phonon interaction, the polarizability may 
be expressed analytically by the Lindhard function. \cite{Mahan2000,giuliani2005quantum}
Conversely, in presence of electron-phonon coupling, the momentum integral 
in Eq.~(\ref{eq:chi0ph2}) needs to be carried out numerically even for the HEG.  
{The Fan-Migdal self-energy in Eq.~\eqref{eq:sigma1} is evaluated by considering 
coupling to a single non-dispersive optical phonon mode 
(Einstein model) with energy $\hbar\omega_{\rm ph} = 100$~meV, 
and an average electron-phonon coupling matrix element 
$\overline g = \langle g^\nu_{\bf kq} \rangle = 100$~meV. 
This value is representative of materials characterized by 
strong coupling to LO phonons such as, for instance, 
TiO$_2$\cite{Verdi2015}, EuO\cite{King2017}, and SrTiO$_{3}$ \cite{Baumberger2016}. } 
We consider the Wigner-Seitz radii $r_s=4$ and $16$, 
which correspond to carrier densities of $2.5 \cdot 10^{22}$ and 
$4 \cdot 10^{20}$~cm$^{-3}$, respectively, which are 
representative of metals and heavily-doped semiconductors.
To quantify the effects of the electron-phonon interaction on the 
dielectric function and on the plasmon dispersion, we inspect 
in the following the loss function  
$L({\bf q},\omega) = {\rm Im} [\varepsilon^{-1}({\bf q},\omega)]$,
which has poles at the energies and momenta resonant with the excitation of 
plasmons and electron-hole pairs.\cite{nozieres1959}
An introduction to the general features of the loss 
function for the 3D HEG can be found elsewhere.\cite{pines1999elementary}

\begin{figure}
   \includegraphics[width=0.48\textwidth]{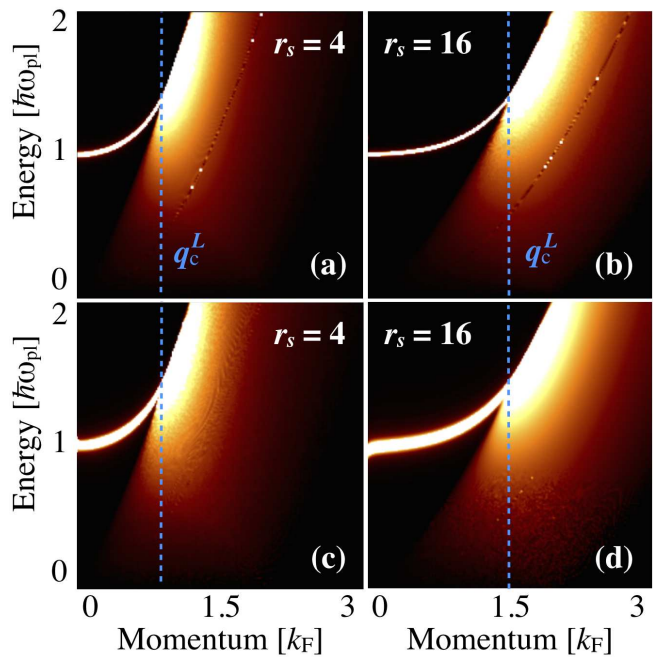} 
\caption{
Loss function of the 3D HEG in the RPA (a)-(b) 
and for the coupled electron-phonon system  (c)-(d) for $r_s = 4$ and $16$.  
The critical momentum $q_{\rm c}^{\rm L}$ (vertical dashed line) marks the onset of 
Landau damping.
Energy is in units of the plasma energy 
($\hbar \omega = 1.2$ 
and $0.15$~eV for $r_s=4$ and 16, respectively), 
whereas momentum is expressed in terms of the Fermi momentum 
($k_F = 0.18$ and $0.045$~\AA$^{-1}$  for $r_s=4$ and 16, respectively).
  }
\label{fig:3D} 
\end{figure}

In Fig.~\ref{fig:3D}, we compare the RPA loss function of the 3D HEG 
with the case of the coupled electron-phonon system. 
In presence of electron-phonon interaction, $L({\bf q},\omega)$  exhibits qualitatively 
similar features to the RPA: there exists a critical momentum cutoff $q^{\rm L}_{\rm c} $ such that 
(i) a sharp plasmon peak is visible at momenta $ q < q^{\rm L}_{\rm c}$
whereas (ii) for larger momenta, $q>q^{\rm L}_{\rm c}$, 
the plasmon energy and momentum become degenerate with those of electron-hole excitations.  
The momentum  cutoff can be expressed as\cite{pines1999elementary} 
$q^{\rm L}_{\rm c} = k_{\rm F}\left[(1+\hbar\omega_{\rm pl}/\epsilon_{\rm F})^{1/2}-1\right]$, 
and its values for $r_s=4$ and 16 are marked in Fig.~\ref{fig:3D} by dashed blue lines, where $k_{\rm F}$ and $\epsilon_{\rm F}$ are 
the Fermi wavevector and the Fermi energy, respectively. 
$\hbar\omega_{\rm pl}=\hbar(4\pi e^2 n /\varepsilon_0m^*)^{\frac{1}{2}}$ is the plasmon energy, which depends on the 
carrier concentration $n$ and amounts to 1.2~eV (0.15~eV) for $r_s = 4$ ($r_s=16$). 
While for  $ q> q^{\rm L}_{\rm c} $ plasmons decay primarily through the excitation 
of electron-hole pairs (Landau damping), 
for $ q < q_{\rm c}^{\rm L} $ phonons constitute the principal plasmon-damping mechanism. 
In this region, the electron-phonon interaction induces a broadening of the plasmon peaks, 
reflecting finite lifetime effects due to the possibility of plasmon damping 
upon phonon emission. 
For both values of $r_s$, the characteristic energy scales of plasmon 
and electron-hole excitations are larger than the phonon energy, and 
phonon-induced damping induces a structureless homogeneous 
broadening of  the loss function. 
The full width at half maximum of the plasmon peak, induced by the coupling to phonons, amounts to 
20 and 5~meV for $r_s=4$ and $16$, respectively. 

A quantitative estimation of the effects of phonons on the characteristic timescales of plasmon damping, 
is provided by the plasmon lifetime, which can be expressed as:\cite{Mahan2000} 
\begin{align}\label{eq:taupl}
\tau_{\bf q} = \hbar^{-1} \left[ \frac{{\rm Im} \,P({\bf q},\omega_{\bf q}^{\rm pl}) }
{ \left[ \partial{\rm Re}\, P({\bf q},\omega) / \partial\omega \right]_{\omega=\omega_{\bf q}^{\rm pl}} } 
\right ]^{-1}.
\end{align}
Through the evaluation of Eq.~\eqref{eq:taupl}, 
we obtain an average lifetime of 130~fs ($750$~fs) 
for $r_s=4$ ($r_s=16$) for momenta $q < q^{\rm L}_{\rm c}$.
The plasmon lifetime drops rapidly to values $\tau_{\bf q}< 1$~fs  for 
$q > q^{\rm L}_{\rm c}$ owing to the effectiveness of the electron-hole scattering mechanism. 

If the coupling to phonons is neglected,  electron-hole scattering 
is the sole decay pathway that plasmons may undergo. Correspondingly, the plasmon 
linewidth is infinitesimal for $q < q^{\rm L}_{\rm c}$, and the lifetime infinite.
The broadening of the plasmon peak in Fig.~\ref{fig:3D}~(a)-(b) 
stems from the imaginary component added to the denominator of 
RPA polarizability and necessary to converge the momentum integrals,\cite{Caruso2016EPJB}
which amount to 5 and 1~meV for $r_s=4$ and $16$, respectively. 

\section{Plasmon-phonon coupling in two dimensions}\label{sec:2D0}

\begin{figure}
   \includegraphics[width=0.48\textwidth]{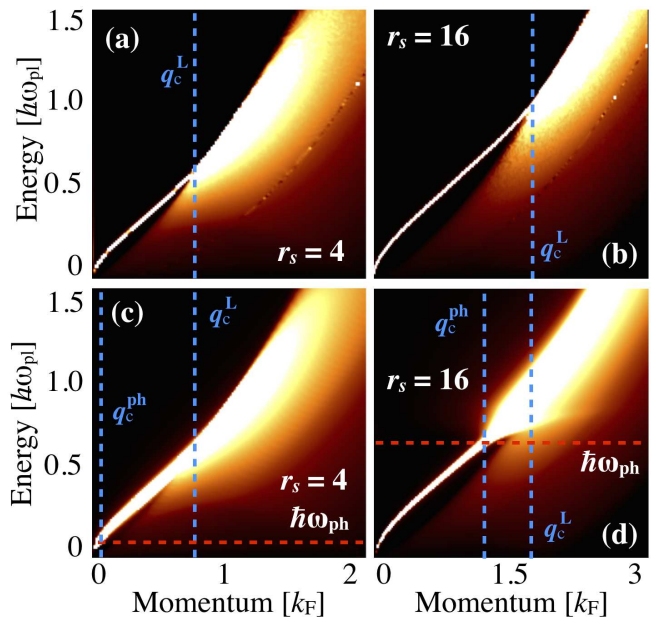} 
\caption{Loss function of the 2D HEG in the RPA (a)-(b) 
and for the coupled electron-phonon system (c)-(d) for $r_s = 4$ and $16$. 
The critical momenta  $q_{\rm c}^{\rm ph}$ and  $q_{\rm c}^{\rm L}$, shown as vertical dashed lines, mark 
the onset of phonon and Landau damping, respectively.
Energy is in units of the plasma energy of the 3D HEG ({\it cf.} caption of Fig.~\ref{fig:3D}).
}
\label{fig:2D}
\end{figure}

We now proceed to discuss general features of plasmon-phonon interaction in the 2D HEG.  
We consider again $r_s = 4$ and 16, which in 2D correspond to carrier concentrations of 
$ 5\cdot10^{15}$  and $1.2\cdot10^{15}$~cm$^{-2}$, respectively. 
The dielectric function is evaluated considering the 2D 
Fourier transform of the Coulomb interaction $v({\bf q}) = 2\pi e^2 / {\bf q } \varepsilon_0$. 
The loss function evaluated in absence of coupling to phonons is reported in  
Fig.~\ref{fig:2D}~(a)-(b), where energies are in units of 
the plasma energy of the 3D HEG.
The lower dimensionality has a striking effect on the energy 
dispersion of plasmons which, in 2D, is proportional to $\sqrt{q}$ 
in the long-wavelength limit.\cite{giuliani2005quantum}
Similarly to the 3D case, there exists a critical momentum $q^{\rm L}_{\rm c}$ such 
that 2D plasmons are undamped for $q < q^{\rm L}_{\rm c}$, whereas  for  $q > q^{\rm L}_{\rm c}$ 
they undergo Landau damping and decay upon excitation of electron-hole pairs. 

When the electron-phonon interaction is switched on, the loss function of the 
2D HEG, illustrated in Fig.~\ref{fig:2D}~(c)-(d), reveals the emergence of more 
complex features as compared to the 3D case. 
For plasmon energies smaller than the phonon energy 
($\hbar\omega_{\rm pl} < \hbar\omega_{\rm ph}$) the coupling between plasmons 
and phonons is forbidden by energy conservation, and the loss function {resembles} 
the case in which electron-phonon coupling is neglected.
Conversely, for $\hbar\omega_{\rm pl} > \hbar\omega_{\rm ph}$ the effects of
electron-phonon coupling on the plasmon dispersions is two-fold: 
(i) the plasmon peak is broadened by finite lifetime effects, 
{(ii) the plasmon energy is renormalized and the corresponding peak is broadened.} 
The combination of (i) and (ii) leads to the formation of a 
{\it kink} in the plasmon dispersion of 2D metals, 
in close analogy to the kink induced in the photoemission 
spectrum of solids by electron-phonon interaction.\cite{Damascelli2003}

In addition to the critical momentum $q_{\rm c}^{\rm L}$ for the onset of Landau damping, 
the coupling to optical phonons, thus, introduces a second critical momentum 
$q_{\rm c}^{\rm ph}$, defined by the condition $\omega^{\rm pl}(q_{\rm c}^{\rm ph})=\omega_{\rm ph}(q_{\rm c}^{\rm ph})$, 
which marks the onset of plasmon decay upon phonon emission. 
An approximate explicit expression for $q_{\rm c}^{\rm ph}$ 
can be obtained by approximating the plasmon dispersion by 
$\omega^{\rm pl}_{\bf q} = (2\pi n e^2 q/m^*\varepsilon_0)^{\frac{1}{2}}$, 
yielding $q_{\rm c}^{\rm ph} = {\omega_{\rm ph}^2m^*\varepsilon_0}/{2\pi n e^2}$. 

The values of $q_{\rm c}^{\rm ph}$ and $q_{\rm c}^{\rm L}$ 
are illustrated in Fig.~\ref{fig:2D}~(c)-(d) as vertical dashed lines. 
In summary, 
for $q < q_{\rm c}^{\rm ph}$ plasmons are undamped; 
for $q_{\rm c}^{\rm ph}< q < q_{\rm c}^{\rm L}$, plasmons may decay upon scattering with phonons; 
for $q > q_{\rm c}^{\rm L}$, both phonons and Landau damping contribute to plasmon dissipation. 
While the coupling to acoustic phonons and electron-electron 
interaction are neglected here, in real solids
these mechanisms are expected to provide additional 
dissipation channels that may lead to plasmon damping 
also in the limit ${\bf q}\rightarrow 0$, and they might 
thus constitute the primary decay mechanism for the 
long-wavelengths plasmons in 2D.\cite{Principi2014,Neilson1991}

{
Finally, we note that, despite the dependence of the numerical results 
of Figs.~\ref{fig:3D} and \ref{fig:2D} on the model parameters $\overline g$ and $\hbar\omega_{\rm E}$,  
this qualitative picture is expected to hold also for parameter values different from those considered above. 
In particular, a decrease in the value of the effective electron-phonon coupling strength
$\overline g$ has the effect of reducing the overall influence of electron-phonon 
interaction on the loss function through a decrease in the broadening of the plasmon peaks.
In the limit of vanishing coupling strength $\overline g \rightarrow 0$, 
the RPA results are recovered.
While the precise value of the phonon energy $\hbar\omega_{\rm E}$ does not
affect significantly the loss function in 3D metals,
in the 2D case $\hbar\omega_{\rm E}$ defines the energy
onset for the scattering between electrons and phonons, marked by
horizontal dashed lines in Fig.~\ref{fig:2D}~(c)-(d). Therefore, different values of the
optical phonon energy are expected to alter the energy window in which the
plasmon dispersion is broadened by lifetime effects.
}

\section{Approximate treatment of electron-phonon interaction }\label{sec:mem}
\begin{figure}
   \includegraphics[width=0.48\textwidth]{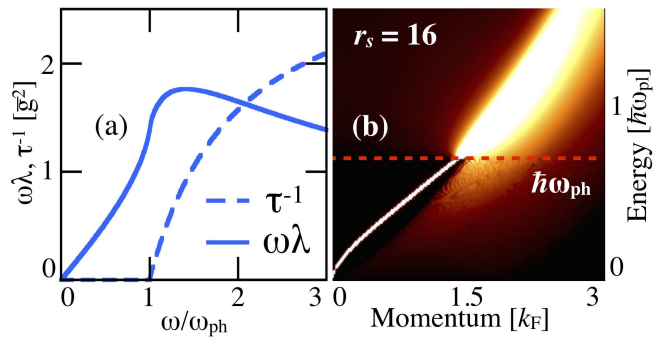} 
\caption{ (a) Dynamical renormalization $\lambda$ and scattering 
time $\tau^{-1}$ for the Einstein model. 
(b) Loss function of the 2D HEG for $r_s=16$ obtained from 
Eq.~\eqref{eq:chi0ph3} using $\lambda$ and $\tau^{-1}$ 
illustrated in panels (a).
}  
\label{fig:mem}
\end{figure}

First-principles calculations of the polarization in the Fan-Migdal 
approximation based on the solution of Eq.~\eqref{eq:chi0ph2} may be possible 
for simple solids, however, the computational 
cost entailed by the evaluation of the electron-hole 
self-energy $\Delta\Sigma({\bf k,q},\omega, \epsilon_{\bf k}/\hbar)$ 
may obstruct the application to systems with more than a few atoms per unit cell.  
To account for the effects of 
electron-phonon interaction in the loss function of crystalline solids, 
it is thus desirable to circumvent the computation of the electron-hole
self-energy through approximations capable of reproducing 
the main features of the loss function at a reduced computational cost.
We follow the approach introduced by Allen\cite{Allen1971} 
and subsequently generalized to first-principles calculations 
in Ref.~\onlinecite{Novko2017}, whereby 
$\Delta\Sigma({\bf k,q},\omega, \epsilon_{\bf k}/\hbar)$
is approximated by its average over the Fermi surface 
in the optical limit (${\bf q}\rightarrow0$).
Within these approximations, the polarization can be expressed as:\cite{Allen1971,Kupcic2017}
\begin{align}\label{eq:chi0ph3}
P({\bf q},\omega) =  2\sum_{\bf k}
\frac{ f_{\bf k} - f_{\bf k+q} }{\hbar\omega[1+\lambda(\omega)] + i/\tau(\omega) + \epsilon_{\bf k} -\epsilon_{\bf k+q}}
\end{align}
where we introduced the dynamical renormalization $\lambda$ and the scattering time $\tau$ functions:
\begin{align}
\tau^{-1}(\omega) &= \frac{2\pi\hbar}{\omega} \int_0^\omega d\omega' (\omega-\omega') \alpha^2F(\omega') \label{eq:tau} \\
\lambda(\omega) &=  -\frac{2}{\omega} \int_0^\infty d\omega' \alpha^2F(\omega') \label{eq:lambda}\\ 
& \,\,\,\,\,\,\,\,\,\,\,\, \,\,\,\,\, \times \left[ \nonumber 
{\rm ln}\left| \frac{\omega-\omega'}{\omega+\omega'}\right| -
\frac{\omega'}{\omega}{\rm ln}\left| \frac{\omega^2-\omega'^2}{\omega'^2}\right| 
\right]
\end{align} 
$\alpha^2F$ is the Eliashberg spectral function and it is defined as: 
\begin{align}
\alpha^2F(\omega) = \frac{1}{2}\sum_{{\bf q}\nu} \omega_{{\bf q}\nu}\lambda_{{\bf q}\nu} \delta (\omega-\omega_{{\bf q}\nu}), 
\end{align}
where the sum extends over all phonon modes $\nu$ and momenta ${\bf q}$. 
{The electron-phonon coupling strength 
is given by:
\begin{align}
\lambda_{{\bf q}\nu} = \frac{1}{{N_{\rm F}\omega_{{\bf q}\nu}}} \sum_{\bf k}|g^\nu({\bf k},{\bf q})|^2 
\delta(\epsilon_{\bf k}-\epsilon_{\rm F})
\delta(\epsilon_{\bf k+q}-\epsilon_{\rm F}),
\nonumber
\end{align}
with $N_{\rm F}$ being the density of states at the Fermi energy. 
If we consider an Einstein model 
for the phonon dispersion with energy $\hbar\omega_{\rm ph}$
and isotropic electron-phonon matrix elements $\bar g$,
the electron-phonon coupling strength and the  Eliashberg spectral function 
simplify to}
$\lambda_{{\bf q}} = \bar g^2 /\omega_{\rm ph}$ and 
$\alpha^2 F(\omega) = {\bar g^2} \delta(\omega-\omega_{\rm ph})/2$. Correspondingly,  
$\tau$ and $\lambda$ can be evaluated analytically from Eqs.~\eqref{eq:tau} and \eqref{eq:lambda}, and yield:
\begin{align}
\tau_{\rm E}^{-1}(\omega) &= \bar g^2\pi\hbar\frac{\omega-\omega_{\rm ph}} {\omega}\theta(\omega-\omega_{\rm ph}). 
\nonumber
\\
\lambda_{\rm E}(\omega) &= - \frac{\bar g^2}{\omega}
\left[ 
{\rm ln}\left| \frac{\omega-\omega_{\rm ph}}{\omega+\omega_{\rm ph}}\right| -
\frac{\omega_{\rm ph}}{\omega}{\rm ln}\left| \frac{\omega^2-\omega_{\rm ph}^2}{\omega_{\rm ph}^2}\right| 
\right].
\nonumber
\end{align}
The resulting scattering rate $\tau_{\rm E}$ and dynamical
renormalization $\lambda_{\rm E}$ for the Einstein model are shown in
Fig.~\ref{fig:mem}~(a).
$\tau^{-1}_{\rm E}$ vanishes for $\omega<\omega_{\rm ph}$,
and it reflects the fact that the transfer of the 
plasmon energy to a phonon is forbidden by energy 
conservation and, correspondingly, no damping due 
to phonons may take place for frequency smaller 
than the phonon frequency. 
At large frequencies $\omega\gg \omega_{\rm ph}$, the damping function
saturates at the value $\tau_\infty^{-1} =\bar g^2\pi$.
These observations are consistent with the results 
of the 3D HEG, shown in Fig.~\ref{fig:2D}, where
the coupling to phonons induces a quasi-homogeneous 
broadening of the loss function for $\hbar \omega \gg \omega_{\rm ph}$.
The loss function obtained by combining Eq.~\eqref{eq:chi0ph3}
with the Einstein model for the scattering time and dynamical renormalization
is exemplified in Fig.~\ref{fig:mem}~(b) for $r_s=16$. 
The model correctly reproduces the emergence
of plasmon damping for $q>{q_{\rm c }^{\rm ph}}$,
and the resulting formation of a kink in the plasmon dispersion. 
Despite the simplicity of the Einstein model, the trend of 
$\tau_{\rm E}$ and $\lambda_E$ is in good qualitative agreement 
with the result of first-principles calculations of Ref.~\onlinecite{Novko2017}.

Overall, these results suggest that the treatment of electron-phonon coupling through 
the effective dynamical-renormalization and scattering-time functions 
suffices to reproduce the main features obtained from the solution of 
the Hedin-Baym equations in the Fan-Migdal approximation, thus, validating earlier 
approximations for the optical conductivity,\cite{Allen1971} as well as 
recent first-principles calculations for doped graphene.\cite{Novko2017}
Additionally, a simplified approach based on the Einstein model may 
also provide a suitable framework to account for electron-phonon 
coupling in first-principles calculations of the dielectric 
properties at a reduced computational cost.
For instance, the $\omega_{\rm ph}$, and $\overline g$ 
can be extracted from first-principles calculations, 
and Eq.~\eqref{eq:chi0ph3} obtained using single-particle 
energies and oscillator strengths 
based on density-functional theory. 

\section{Conclusions}\label{sec:conc}
We have investigated the influence of the electron-phonon interaction 
on the dissipation of plasmons in metals based on the 
solution of the Hedin-Baym equations for the dielectric 
function in the Fan-Migdal approximation. 
We have reported calculations of the loss function of the HEG in 2D and 3D 
and discussed the general features induced by electron-phonon coupling in the 
plasmon dispersion for carrier densities representative of metals and heavily-doped semiconductors. 
The coupling to phonons in 3D leads primarily to a broadening 
of the plasmon peak and to finite-lifetime effects.  
In addition to these signatures of plasmon-phonon coupling, 
our loss-function calculation for 2D systems also indicate 
the emergence of a kink in the plasmon dispersion -- in close analogy 
to the kinks in photoemission spectroscopy -- which arises from the
coupling to a longitudinal optical phonon mode.
In conclusion, these results call for further work to investigate 
compensation effects arising from the inclusion of vertex corrections,\cite{Bechstedt1997} 
the interplay of the electron-electron and electron-phonon interaction on the plasmon dispersion, 
as well as the temperature dependence of plasmon damping.

\acknowledgments
Discussions with Pierluigi Cudazzo and Lucia Reining are gratefully acknowledged. 
{FC and CD gratefully acknowledge financial support from the 
German Science Foundation (DFG) through the Collaborative Research 
Center HIOS (SFB 951).}


%

\end{document}